\documentclass[12pt,oneside,english]{amsart}
\usepackage[T1]{fontenc}
\usepackage[utf8]{inputenc}
\usepackage[letterpaper]{geometry}
\usepackage{booktabs}
\usepackage{amstext}
\usepackage{amsthm}
\usepackage{amssymb}
\usepackage{setspace}
\usepackage{subcaption}
\usepackage{caption}
\usepackage{float}
\usepackage [displaymath,mathlines]{lineno}
\usepackage{xcolor}
\usepackage{graphicx}
\usepackage{pdfpages}
\usepackage{verbatim}
\usepackage{multirow}
\newcommand\independent{\protect\mathpalette{\protect\independenT}{\perp}}
\def\independenT#1#2{\mathrel{\rlap{$#1#2$}\mkern2mu{#1#2}}}

\makeatletter

\numberwithin{equation}{section}
\numberwithin{figure}{section}
\theoremstyle{plain}

\usepackage{chngcntr}
\usepackage{apptools}
\AtAppendix{\counterwithin{lemma}{section}}

\theoremstyle{plain}

\theoremstyle{plain}

\numberwithin{equation}{section}

\usepackage[within=none]{caption}

\usepackage{babel}
\usepackage{amsmath}
\usepackage{mathtools}
\usepackage{bbm}
\usepackage{relsize}
\usepackage{bm}

\usepackage{algpseudocode}

\usepackage[ruled]{algorithm2e}

\usepackage{array}
\newcommand{\mockalph}[1]{}

\makeatother

\providecommand{\theoremname}{Theorem}

\begin{document}
\title{Bayesian outcome selection modelling}
\author{Khue-Dung Dang$^{1\ast}$, Louise M. Ryan$^{2,3}$, Richard J. Cook$^{4}$, Tugba Akkaya-Hocagil$^{4}$, Sandra W. Jacobson$^{5}$ And Joseph L. Jacobson$^{5}$ }

\singlespacing
	\thanks{\small{$^1$ \textit{School of Mathematics and Statistics, University of Melbourne, Melbourne, Australia} \\
	$^2$ \textit{School of Mathematical and Physical Sciences, University of Technology Sydney, Sydney, Australia} \\  $^3$ \textit{Australian Research Council Centre of Excellence for Mathematical and Statistical Frontiers, Australia}\\ $^4$ \textit{Department of Statistics and Actuarial Science, University of Waterloo, Waterloo, Canada} \\$^5$ \textit{Department of Psychiatry and Behavioral Neurosciences, Wayne State University School of Medicine, Detroit, Michigan, USA}\\ $^\ast$ Corresponding author: kd.dang@unimelb.edu.au}}
	\doublespacing
	\maketitle
	\begin{abstract}
	Psychiatric and social epidemiology often involves assessing the effects of environmental exposure on outcomes that are difficult to measure directly.
	To address this problem, it is common to measure outcomes using a comprehensive battery of different tests thought to be related to a common, underlying construct of interest. In the application that motivates our work, for example, researchers wanted to assess the impact of \textit{in utero} alcohol exposure on child cognition and neuropsychological development, which were evaluated using a range of different tests. Statistical analysis of the resulting multiple outcomes data can be challenging, not only because of the need to account for the correlation between outcomes measured on the same individual, but because it is often unclear, \textit{a priori},  which outcomes are impacted by the exposure under study. While researchers will generally have some hypotheses about which outcomes are important, a framework is needed to help identify outcomes that are sensitive to the exposure and to quantify the associated treatment or exposure effects of interest. We propose such a framework using a modification of stochastic search variable selection (SSVS), a popular Bayesian variable selection model and use it to quantify an overall effect of the exposure on the affected outcomes. We investigate the performance of the method via simulation and illustrate its application to data from a study involving the effects of prenatal alcohol exposure on child cognition.  
	
	
\end{abstract}

\section{Introduction}

Multiple outcomes are routinely collected in psychological and social epidemiology studies in which participants are assessed on a comprehensive battery of tests or tasks designed to measure psychological, neurological or cognitive outcomes that are difficult to measure directly. A large number of outcomes are often collected, so it can be challenging to decide which outcomes to include in the analysis. Scientists typically rely on previous studies, in combination with expert knowledge to select the outcomes on which to focus. Until recently, no statistical framework for identifying outcomes that are sensitive to an exposure has been developed, nor has such a framework been developed to quantify the magnitude of effects. 

Our work is motivated by a cohort study aimed at providing insights into the precise impact of prenatal alcohol exposure (PAE) on child development. Numerous studies have shown that high levels of PAE can result in a distinct pattern of craniofacial anomalies, growth restriction, and cognitive and behavioral deficits, a condition known as fetal alcohol syndrome (FAS) \cite{hoyme2005practical, hoyme2016updated}, the most severe of a continuum of fetal alcohol syndrome disorders (FASD) \cite{carter2016fetal, jacobson2004maternal,jacobson2008impaired,mattson2019fetal}. Alternatively, some individuals with PAE exhibit cognitive and/or behavioural impairment without the characteristic craniofacial dysmorphology and/or growth restriction, a disorder known as alcohol-related neurodevelopmental disorder (ARND). It is also critical to understand the more subtle effects of low-dose exposures. In this longitudinal study, funded by the US National Institutes of Health, expectant mothers were interviewed about their drinking habits during the pregnancy, and their children were followed throughout childhood and adolescence, many of them until they were 20 years of age \cite{jacobson2004maternal,jacobson1993teratogenic}. At every follow-up visit, children were assessed using a broad range of cognitive and neuro-developmental tests. Each of the administered tests could be classified as relevant to one of several different domains including cognition, executive function, and behaviour, among others. Previous neurocognitive studies have suggested that the impact of PAE on all of these domains will not be the same, given that alcohol may have a stronger effect on certain parts of the brain, while other areas may be relatively unaffected or spared, depending on the timing, genetic vulnerability, and ethnic or racial group of the exposure \cite{jacobson2004maternal,jacobson1999drinking, jacobson2002effects}. Recent analyses by our group made use of expert knowledge to select outcomes for analysis and simply assumed that each had been affected by PAE to some extent \cite{jacobson2021effects}. However, a more sophisticated modelling framework is needed to discriminate between  outcomes that are strongly or weakly affected by PAE. In this paper we develop and evaluate a method which address this need. 

There is a rich literature on statistical methods for the analysis of multiple outcomes data. A simple approach is to analyze each outcome separately, but this approach can result in reduced power due to the need to adjust for multiple comparisons \cite{lefkopoulou1993global}. Structural equation models (SEM) can also be used to model correlated outcomes by introducing latent variables and treating the outcomes as manifestations of the latent variables \cite{budtz2002estimation,dunson2000bayesian,sanchez2005structural}. The estimated factor loadings can provide some insight regarding whether or not the various outcomes are representative of the hidden factors. However interpreting regression coefficients characterizing the relationship between the exposure and the latent factor can be problematic and inferences can be very sensitive to model misspecification \cite{sammel2002effects}. Meta\mbox{-}analysis is another popular approach to synthesis of multiple outcomes data, but relatively little work has been carried out for dealing with highly correlated outcomes in observational settings. \cite{berkey1998meta,van2015meta,ryan2008combining}.  Meta-analysis, adjusted for between\mbox{-}outcome correlations, was the methods used in the previous paper that we published on this topic \cite{jacobson2021effects}.  Generalized estimating equations \cite{liang1986longitudinal} have also been used to analyze multiple outcome data, with working covariance matrices specified to accommodate correlations across outcomes, since the repeated observations on each individual can be viewed as a special type of clustered. Generalized linear mixed models offer another framework to model the effect of exposure on
multiple outcomes \cite{sammel1999multivariate,thurston2009bayesian}. 

A limitation of the available statistical methods for analyzing multiple outcomes data is that researchers must specify the outcomes to be included in the analysis. This is usually done using expert knowledge or following some gatekeeping procedure to select the subset of affected outcomes (see, for example, \cite{turk2008analyzing}), but this is challenging when outcomes are high dimensional or when expert knowledge does not provide strong \textit{a priori} guidance. Moreover, using exploratory data analysis to guide the decision\mbox{-}making increases the risk of distorting inferences due to multiple comparisons. There is a clear need for a principled statistical approach for identification of relevant outcomes on which to model the exposure effects, whilst accounting for the correlation among the outcomes. 

As we will show presently, the challenge of identifying which of many observed outcomes are sensitive to an exposure can be reframed as a variable selection problem. 
Variable selection is typically carried out to choose a subset of candidate predictors that together explain most of the variation in a single response variable. The variable selection literature has a long history, from earlier frequentist approaches such as ``best subset'' regression, model selection based on Akaike/Bayesian information criterion \cite{akaike1998information,schwarz1978estimating}, backward and forward stepwise regression, to the more recent Bayesian methods that involve a wide range of ``slab-and-spike'' or shrinkage priors (see Hastie et al. \cite{hastie2020best}, O'Hara and Sillanp{\"a}{\"a} \cite{o2009review} and Van Erp et al. \cite{van2019shrinkage} for reviews and  discussions of some of these popular approaches). Despite this large literature, there is no method incorporate selection of key outcomes from a number of potentially correlated outcomes, with a view of identifying those that are sensitive to exposure. 


In this paper, we first show how the multiple outcomes modelling problem can be reframed as one of variable selection. We adopt a Bayesian approach to analyze outcomes and identify those that are strongly affected by the exposure. The model is motivated by the popular stochastic search variable selection (SSVS) method for variable selection, though we extend the SSVS prior to allow estimation of a mean effect among the sensitive outcomes. We propose random effect model to account for the correlation among the outcomes of each individual.

The paper is organized as follows: following this introduction, in Section \ref{sec:model} we present the basic model and discuss the associated computing approach. In two simulation studies in Section \ref{sec:simExample} we assess the performance of our method in comparison to other variable selection models. In Section \ref{sec:RealApplication} we use the model to analyze the data from our motivating application related to the effect of \textit{in utero} alcohol exposure on different measures of child cognition. In Section \ref{sec:Conclusion} we present the conclusions of the paper.

\section{Methodology} \label{sec:model}
\subsection{Addressing the outcome selection problem}
\label{sec:model_classical}
Suppose we observe $K$ continuous outcomes for each of the $n$ individuals. The outcomes will typically be correlated because they are measures from the same individual, though they may be of different scales and nature. For example, the outcomes measure different domains of a person's cognition function. Therefore, exposure effects are not expected to be exactly the same across affected outcomes but vary around a mean level $\mu$, which we identify as the parameter of interest. For each individual, we observe the exposure variable $x$ and use $z$ to denote other observed predictor variables that can be added to the model. In the application to our motivating study in Section \ref{sec:RealApplication}, $z$ is a propensity score computed for each individual to adjust for confounders.

We now show how to express our multiple outcomes data as panel data in long format. Consider a sample of $n$ independent individual labeled $j = 1,\dots,n$, where each individual provides $K$ outcomes labeled $p = 1,\dots, K$. Suppose we stack the observations of all $n$ individuals together, this gives us a data set with $nK$ observations in total. 
Row $i$ of this data set records the observed outcome corresponds to individual $j[i]$ and outcome $p[i]$. 

\begin{figure}
	\centering
	\includegraphics[width=0.7\linewidth]{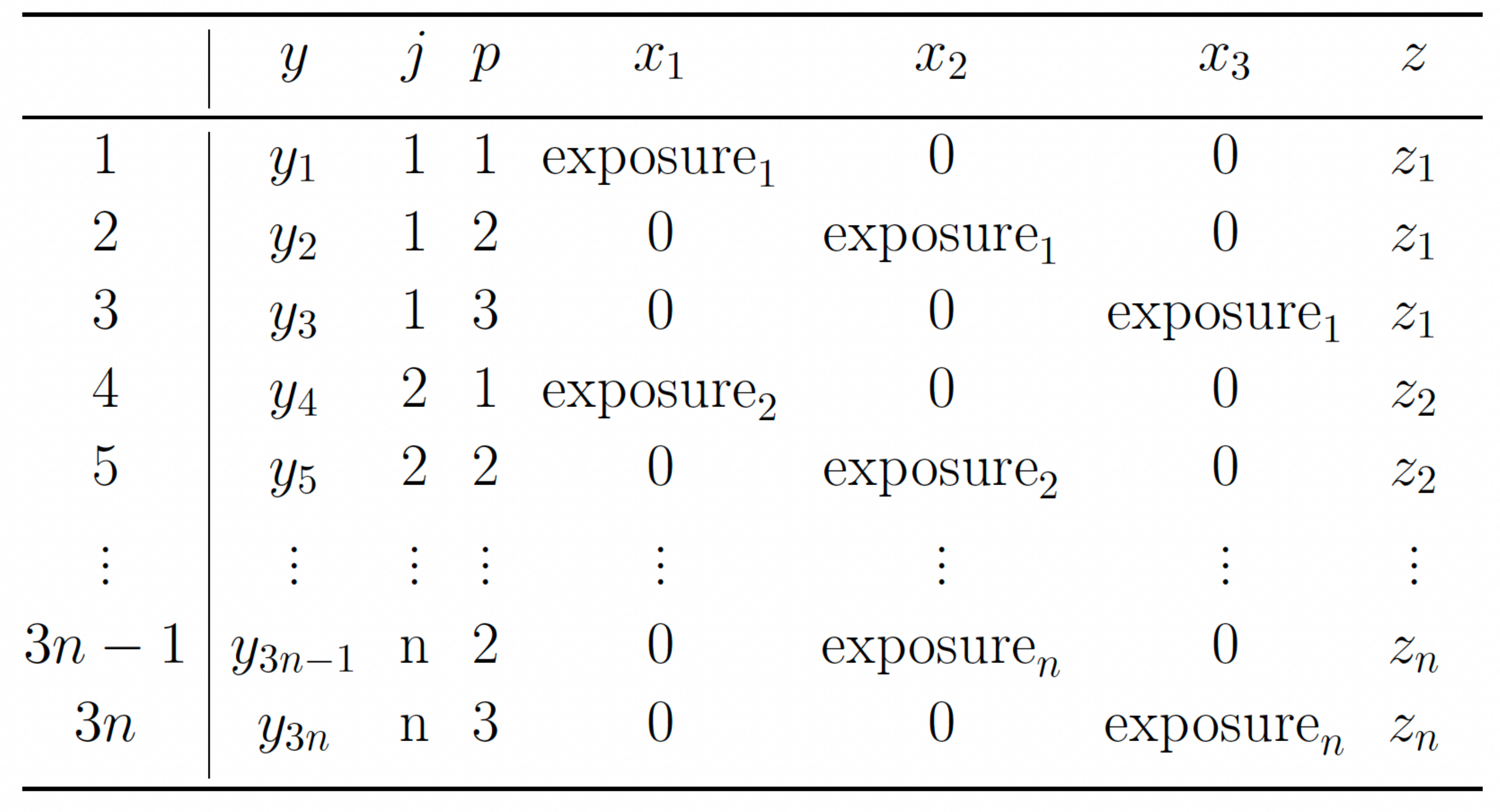}
	\caption{Illustration of a data table with $n$ individuals and $K = 3$ outcome variables.}
	\label{fig:longdata}
\end{figure}

To represent the multiple outcome problem as a multiple predictors problem, we first define a new set of covariates $x_1, \dots, x_K$, $$ x_{ik} = \text{exposure}_{j[i]} \boldsymbol{1}(p[i] = k),$$ for $i = 1,\dots, nK$, individual $j = 1,\dots, n$ and outcome $p = 1,\dots, K$. These are the interaction terms between the exposure level and the dummy variables indicating the outcome variables. An example of the dataframe format and how to map from the multiple outcome format to the stacked format for $n$ individuals and $K = 3$ outcome variables is presented in Figure \ref{fig:longdata}.

We then model the $y_i$ using a linear mixed model:

\begin{equation}
   \label{eq:model_simple}
   y_{i} = \nu_{p[i]} + \alpha_{j[i]} + \sum_{k=1}^K \beta_k x_{ik} +  \gamma_{p[i]} z_{j[i]}  + \epsilon_{i},
\end{equation}
for $i = 1,\dots, nK$, individual $j = 1,\dots, n$ and outcome $p = 1,\dots, K$. The error terms $\epsilon_{i}$ are independent and normally distributed, $\epsilon_{i}\sim N(0,\sigma^2_{p[i]})$ and the random effect $\alpha_{j[i]} \sim  N(0,\sigma_r^2)$ accounts for the within-individual correlation and $\alpha_j \independent \alpha_{j^\prime}$ for $j \neq j^\prime$.

The parameters $\nu_{p}$ and $\gamma_p$ are outcome-specific intercepts and coefficients for $z$, and the coefficients $\beta_k$ represents the exposure effect on outcome $k$, assumed to vary around a common value $\mu$:
$$ \beta_k  \sim N(\mu,\tau^2),$$ for $k = 1,\dots,K$. 

This "trick" of expressing the multiple outcomes problem as a repeated measures problem has been widely used in the literature, making it straightforward to then analyze multiple outcomes using standard mixed modelling or GEE software \cite{lefkopoulou1989analysis}. Now we have a mixed model with both varying intercepts and varying coefficients. We will perform variable selection on the $K$ covariates $x_1,\dots, x_K$.
 

If we assume that all outcomes are associated in a similar way with the exposure or treatment of interest, with effects varying around a mean level $\mu$, then we can assign appropriate priors for $\mu$ and $\tau$ and estimate the model using a Bayesian estimation procedure. In this new outcome selection framework, we assume that some of the outcomes are not relevant to the analysis. This implies $\beta_k = 0$ for some $k$.  We then use variable selection priors to analyze $\beta_k$. 

Over the past decade or so, variable selection methodologies have been 
extended to the random effect context, see for example \cite{bondell2010joint,fan2012variable} and \cite{yang2020bayesian}. For our model, because we do not need to perform variable selection for the random effects, it is straightforward to use existing techniques for the independent predictors $x_k$. This means that we can potentially use any of a variety of sparsity priors, such as stochastic search variable selection \cite{george1993variable}, Bayesian LASSO \cite{figueiredo2003adaptive,park2008bayesian} and the horseshoe prior \cite{carvalho2009handling} for outcome selection. However, because we are interested in selecting the subset of affected outcomes variables and quantifying the mean exposure effect on these variables, in this study we focus on the use of the stochastic search variable selection method.  We discuss this further in the next section.


 \subsection{Stochastic search variable selection}
There is a large literature on Bayesian variable selection methods, however in this paper we only discuss the ``slab and spike'' type of priors, as they are suitable for our problem of identifying sensitive outcomes from a large number of outcomes. Methods that compare models by Bayes Factor \cite{kass1995bayes} or criteria such as DIC \cite{spiegelhalter2002bayesian} or WAIC \cite{watanabe2010asymptotic} require fitting all candidate models and hence only applicable when comparing a small number of models. Therefore, they are not suitable for the outcome selection problem.

Methods involving a ``slab and spike'' prior can be divided into two categories: Methods that specify a prior that approximate the ``slab and spike'' shape for the coefficients $\beta_k$; and methods that use latent indicator variables that indicate whether a covariate is included in the model. Shrinkage priors such as the Bayesian LASSO \cite{figueiredo2003adaptive,park2008bayesian} and the horseshoe prior \cite{carvalho2009handling} belong to the first category. The implementation of these methods is straightforward and they have extensive use in the recent years, however it is not possible to modify these priors to incorporate a common mean of the non-zero coefficients. 

The second category defines a latent variable $I_k$ that indicates whether a coefficient $\beta_k$ is non-zero. In the approaches proposed by \cite{kuo1998variable} and \cite{dellaportas2002bayesian}, a coefficient $\beta_k$ is set to 0 if $I_k = 0$. These approaches are harder to tune to ensure the iterates of $I_k$ do not get stuck at $0$ or $1$. Our method is motivated by the stochastic search variable selection (SSVS) method \cite{george1993variable}, which defines a mixture prior for $\beta_k$ instead: Let $I_k$ be a latent indicator variable, with $I_k = 1$ means covariate $k$ is included in the model, $I_k = 0$  means it is not. The indicator affects the prior of $\beta_k$, so we can define a joint prior for $(I_k,\beta_k)$ as $$p(I_k,\beta_k) = p(\beta_k|I_k)p(I_k).$$ Conditioning on $I_k$, the prior of $\beta_k$ is
 $$p(\beta_k|I_k) = (1-I_k)N(0,g_1) + I_kN(0,\tau^2). $$


 The tuning parameters $g_1$ and $\tau^2$ specify the scale of the conditional prior of $\beta_k$ given $I_k$ and are typically data dependent: they should take into account the posterior distribution of the $\beta_k$ and hence the size of the data \cite{george1993variable}. Ideally, $g_1$ should be small enough such that if $\beta_k \sim N(0,g_1)$, then $\beta_k$ can be safely replaced by $0$. On the other hand, $\tau$ must be large enough such that non-zero $\beta_k$ can be included in the model. These parameters can be fixed or estimated as the model's parameters \cite{george1993variable}. Another variation of SSVS is when $g_1$ being $\tau$ multiplied by a constant \cite{meuwissen2004mapping}.  

 For our outcome selection framework, we propose to modify the SSVS prior to incorporate the mean exposure effect $\mu$ on the sensitive outcomes. Conditioning on $I_k$, we now have a mixture prior for $\beta_k$ 
 \begin{align}\label{eq:SSVSprior}
 p(\beta_k|I_k) = (1-I_k)N(0,g_1) + I_kN(\mu ,\tau^2).
 \end{align}
 To improve the performance of the model, we follow \cite{meuwissen2004mapping} and modify the prior in \eqref{eq:SSVSprior} to 
 \begin{align}\label{eq:SSVSpriorRatio}
 p(\beta_k|I_k) = (1-I_k)N(0,\tau^2/c) + I_kN(\mu ,\tau^2).
 \end{align}
 
 The constant $c$ can be chosen by trial-and-error to ensure good separation between the `in' and `out' variables. The standard deviations $\tau$, $\sigma_k$ and $\sigma_r$ are assigned log-normal priors in our simulation examples and real data application. 
 
  Note that the posterior mean of $I_k$ will be the posterior probability that outcome $k$ is included in the model, hence it will be important in terms of interpreting the results of our model fit. The prior, $p(I_k=1)$, can simply be a categorical distribution with a fixed probability parameter. This prior probability may be different across outcomes, based on the experts' knowledge, or fixed at 0.5 so that the prior is non-informative. We can also treat $p(I_k=1)$  as a parameter to be estimated \cite{o2009review}.
 

\section{Simulation examples} \label{sec:simExample}
We now demonstrate the performance of the method in two scenarios. In the first scenario, we consider the performance of the model when the data are generated from the correct model in \eqref{eq:model_simple}. Then in Section \ref{subsec:simSEMdata} we assess method's performance in a more challenging scenario. 

The aim of the simulation studies is to assess the performance of our prior for outcome selection for different effect sizes. In both exercises, we set the number of outcomes to $K = 20$ and a moderate sample size $n = 100$. We examine the performance of the proposed model with different numbers of relevant outcomes $K_1 = 5$, $10$, $15$. 

For each setting we simulated 10 data sets, and for each data set we fit our model in \eqref{eq:model_simple} using the prior in \eqref{eq:SSVSpriorRatio} with $c= 100$. We examine both versions of SSVS: our proposed model in which $\mu$ is a parameter and the standard SSVS prior where $\mu = 0$. The prior probability of $I_k$ is $p(I_k=1) = 0.5$ for all outcome $k$. For comparison, we also fit the model where $\beta_k$ are assumed a hierarchical prior $\beta_k \sim N(\mu,\tau^2)$ with unknown $\mu$ and $\tau^2$. We also fit the model that only uses the correct relevant outcomes, assuming $\beta_k\sim N(\mu,\tau^2)$. We call this the ``subset model''. The result of the subset model is treated as the ``standard'' because it is the model that uses the correct set of outcomes. 

The rest of the parameters are assigned fairly flat priors. We use a normal $N(0,100)$ prior for $\mu$, log-normal$(0,10)$ for $\sigma_r$ and $\sigma_k$, $k = 1,\dots, K$. The prior for $\nu_k$ and $\gamma_k$ are normal $N(0,1000)$ and normal $N(0,100)$, respectively. In all models, $\tau$ is assigned a log-normal$(0,1)$ prior. 

For each simulation we record the number of outcomes identified as relevant, the number of correctly identified outcomes, the number of false positive and the estimated $\mu$. An outcome is classified as ``relevant'' if the posterior mean of the corresponding $I_k$ is greater than 0.5. The results presented here is the average over the 10 simulations for each setting. In each simulation, we have the exogenous variable $z_i \sim N(0,1)$ and $x_i|z_i \sim  \boldsymbol{1}(z_i<0)N(0,0.5^2) + (1-\boldsymbol{1}(z_i<0))N(1,1)$.

The SSVS models are fitted using the software JAGS \cite{plummer2003jags} and the other models are implemented with STAN \cite{carpenter2017stan}.

\subsection{Simulation study 1: Performance under correct model specification}\label{subsec:simREdata}
In the first simulation example, we generated data from the model \eqref{eq:model_simple}:
\begin{align*}
y_{ik} &= \nu_k + \beta_kx_i + \alpha_i + \gamma_k z_i + \epsilon_{ik}; \quad \epsilon_{ik} \sim N(0,\sigma_k^2).
\end{align*} 
This is the same as the model that is the basis for our proposed method in Section \ref{sec:model_classical}.
The data were generated as follows: The intercepts were generated from $N(0,1)$ and the standard deviations $\sigma_k^2$ are generated from $N(1.5,0.3)$. The coefficients $\beta_k$ corresponding to the relevant outcomes were generated from $N(\mu,0.01\mu^2)$ to ensure that the $\beta_k$ are scattered closely enough around $\mu$. We used two different values for the mean common effect $\mu$: $\mu = -0.1$ and $\mu = -3$. The first setting represents a situation when the effect is weak with small data, so that the posterior standard deviation is large. In this case it would be difficult for the model to decide whether a $\beta_k$ is 0 or not. The second setting mimics the situation in which the effect is stronger, and the selection method is expected to work better.


The performance of the SSVS algorithm in detecting the affected outcomes for different $\mu$ and $K_1$ is presented in the top 6 rows of Table \ref{tab:simLong}. The result shows that the original and our modified SSVS algorithms have very similar performance, but both do not do well in detecting the affected outcomes when $\mu$ is small. This is expected as the overall effect $\mu$ is small, so some of the relevant $\beta_k$ would be close to 0. Because here we used uninformative priors for $I_k$, the algorithm will keep switching between stage $I_k=0$ and $I_k= 1$ for these outcomes. This is similar to the phenomenon observed by O'Hara and Sillanp{\"a}{\"a} in \cite{o2009review}, where the posterior probabilities of $I_k$ are close to $0.5$ and some outcomes are classified incorrectly by chance. 

Table \ref{tab:simMSE} shows the mean squared errors of estimating the individual coefficient $\beta_k$:
\begin{equation}\label{eq:MSE}
    \text{MSE} = \frac{1}{K}\sum_{k =1}^K(\widehat{\beta}_k-\beta_k)^2,
\end{equation}
where we take the estimate $\widehat{\beta}_k$ to be the posterior mean of $\beta_k|I_k = 1$ if the posterior mean of $I_k$ is greater than 0.5; otherwise we set $\widehat{\beta}_k=0$. For both large and small values of $\mu$, the SSVS priors provide more accurate estimates of $\beta_k$ in terms of MSE, compared to the model without variable selection. 

Lastly, Table \ref{tab:simMu} shows the estimated of $\mu$, averaged over 10 simulations, by different priors. Table \ref{tab:simMu} shows that our modified SSVS can provide estimates of $\mu$ that are closer to the result from the subset model, especially for large $\mu$. However, when the effect is weak, the model is not able to estimate $\mu$ accurately because it fails to identify the correct set of sensitive outcomes. 

This simulation example shows that SSVS priors can provide accurate estimates of the coefficients, and accurately identify the affected outcomes and estimate the mean effect when $\mu$ is far from 0. However, it may require more informative priors for $I_k$ and better tuning to capture small effects accurately. 

\subsection{Simulation study 2: Performance under model misspecification}\label{subsec:simSEMdata}
In this example, we attempt to fit the model to less ideal data, which have a more complicated correlation structure.  The data were generated according to a confirmatory factor analysis as follows:
\begin{align} \label{eq:simModel}
\bm{y}_i &= \bm{\nu} + \bm{\Lambda}\begin{bmatrix}
\eta_{1,i}\\ \eta_{2,i}
\end{bmatrix} + \bm{e}_i ,\quad \bm{e}_i\sim N(0,\Psi)\\
\eta_{1,i} &= \omega x_i + 0.1 z_i + u_i,\quad u_i \sim N(0,1) \\
\eta_{2,i} &\sim N(0,1).
\end{align}
Here only the first factor $\eta_{1}$ is linked to the exposure $x$ while the second factor does not. In this experiment, although we are not interested in fitting factor analysis models, we find them to be a useful strategy for generating data with correlated outcomes, but where only some of them are influenced by the exposure $x$. In the example $\bm{\Lambda}$ will be a $K\times2$ matrix where $\Lambda[k,1] = 0$ for $k > K_1$, for $K_1$ being the number of relevant outcomes that load into $\eta_{1,i}$. When generating the data, the elements of $\bm{\Lambda}$ correspond to $\eta_{1}$ and $\eta_{2}$ were generated from $N(1,0.1)$ and $N(1,0.2)$ respectively. The intercepts were randomly sampled from $N(0,100)$. The covariance matrix $\Psi$ of the error term $\bm{e}_i$ is diagonal.


Here we still examine the ability of SSVS to correctly capture the relevant outcomes in various settings and compare it to other approaches. We generated data with $n =100$ individuals and total number of outcomes $K = 20$ from the two-factor model as outlined at the beginning of this section. We test our outcome selection priors for $\omega = -0.5$ and $\omega = -3$ while varying the number of relevant outcomes in $K_1 = 5,10,15$.

The last 6 rows of Table \ref{tab:simLong} record the performance of different priors in identifying the relevant outcomes in the two scenarios. An outcome is classified as ``relevant'' by the fitted model if the posterior mean of the corresponding $I_k$ is greater than 0.5. Table \ref{tab:simLong} shows that, in this non-ideal situation, the proposed model may classify an affected outcome as irrelevant, or vice versa when the effect is weak. On the other hand, when the effect $\mu$ is strong, most of the time the model correctly identifies the relevant outcomes. Our proposed prior where $\mu$ is treated as a parameter seems to perform slightly better than the original SSVS where $\mu= 0$.

The bottom part of Table \ref{tab:simMSE} presents the MSE in estimating the coefficients, and we take $\beta_k = \bm{\Lambda}_{k,1}\times \omega$. Similar to the previous experiment, the SSVS priors give more accurate estimates of $\beta_k$ for $\omega = -3$, and have similar accuracy with the model without variable selection when the overall effect is small.  

Table \ref{tab:simMu2} shows the mean effect $\mu$ estimated by different methods. Note that the $\omega$ we used to generate the data is not equal to the true mean effect $\mu$. But the implied true value of $\mu$ can be computed by multiplying $\omega$ with the average of the non-zero loadings corresponding to the first factor. We provide these values in the third column of Table \ref{tab:simMu2}. Similar to the previous experiment, our proposed model does not perform well in the first scenario where $\omega= -0.5$. In both cases, inclusion of the irrelevant outcomes results in a smaller estimate of $\mu$. Not surprisingly, our modified SSVS prior shows similar results with the subset model and their estimates are close to the true values when $\omega = -3$. On the other hand, all priors being considered do not give accurate estimates of mean effect for small $\omega$. This is consistent with our observation in Section \ref{subsec:simREdata}.




\section{Application: Effect of prenatal alcohol exposure on children in Detroit, Michigan} \label{sec:RealApplication}
In this section we return to the application that motivated our work. Specifically, we apply our model to select outcomes from a longitudinal cohort study conducted in Detroit, Michigan, whose aim was to investigate the long-term effects of PAE on a child’s cognitive and behavioral function. In this longitudinal study, the mothers were interviewed prenatally about their alcohol consumption during pregnancy, and the children were followed throughout childhood, many of them up until they were 20 years of age. At the time of each follow up visit, each child was assessed using a large number of cognitive tests. PAE is computed based on the mother’s average daily dose of absolute alcohol consumed (in ounces) during pregnancy (AA/day). Because the distribution of alcohol exposure is positively skewed with a minimum level 0, we compute log(AA/day + 1) and use this as the measure of PAE in the analysis. 

The Detroit study collected a large number of variables reflecting responses on various neuro-cognitive tests and behavioral outcomes assessed on the children throughout childhood. To illustrate our methodology, we focus on a set of 14 outcomes collected when the children were approximately 7 years of age.  The first 8 outcomes come from the Achenbach Child Behavior Checklist (CBCL) and Teacher’s Report Form (TRF) at age 7 \cite{achenbach1991manual}. The CBCL is a checklist completed by the parents and designed to detect emotional and behavioral problems in children and adolescents, while the TRF represents the child’s principal teacher’s report of the similar. These assessments include the child’s internalizing and externalizing behaviors, and social and attention problems. The remaining 6 outcomes correspond to the results of various cognitive and neuro-developmental tests related to IQ assessed on the Wechsler Intelligence Scales for Children--III \cite{wechsler1991manual}, academic achievement in reading and arithmetic, learning and memory abilities and executive function. We have recently reported that, of these 14 outcomes, the first 8 are relatively less affected by PAE, whereas the last 6 are more sensitive to alcohol exposure \cite{jacobson2021effects}.

 
 \subsection{Model and setup}
 
 We fit the model \eqref{eq:model_simple} with the prior in \eqref{eq:SSVSpriorRatio} to the data set. To adjust for confounders associated with both alcohol exposure and cognitive function we add a propensity score $z$ which was computed beforehand. For details on the covariates included in the propensity score and how it was constructed we refer readers to \cite{hocagil2020pscore}.
 Before running the analysis, we rescaled  all outcomes to mean 0 and variance 1.

We fit our proposed model with a few different settings. We start with an uninformative prior for the indicator $I_k$ and set $p(I_k =1) = 0.5$ for all $k$. We use the prior in \eqref{eq:SSVSpriorRatio} with $c = 100$ for $\beta_k$ where $\tau$ is assigned a log-normal$(0,1)$ prior. We also choose a normal $N(0,1)$ prior for $\mu$. As a comparison, we also try the prior in \eqref{eq:SSVSprior} where we fix $g_1 = 0.2^2$ and a shrinkage prior. For the shrinkage prior, we simply follow \cite{figueiredo2003adaptive} and assign a $\text{Laplace}(0,1)$ prior for the $\beta_k$. We also attempt the horseshoe prior \cite{carvalho2009handling} for $\beta_k$ but the MCMC has convergence issue and hence the result is not presented here. 

To assess how sensitive the result is to the prior probability $p(I_k =1)$, we also fit the model with a more informative set of $p(I_k =1)$,
$$ p(I_k =1)= (0.5, 0.5,0.2,0.5,0.8, 0.8, 0.2, 0.8, 0.5, 0.5, 0.8, 0.8,0.8, 0.5).$$ 
These prior probabilities were chosen by utilizing expertise knowledge and set the probability of the outcomes that are known to be relevant to be closer to 1. In practice, more informative priors may help the MCMC to have better mixing. 


We also fit the model \eqref{eq:model_simple} to only those outcomes chosen by our SSVS model with a hierarchical prior $\beta \sim N(\mu, \tau^2)$. We call this model the ``subset'' model. Similar to in the simulation studies, we will compare the estimates of $\beta_k$ and $\mu$ from this reduced model with our approach.

For the rest of the parameters, we choose relatively non-informative priors. We assign a normal $N(0,1)$ prior for $\mu$, log-normal$(0,10)$ for $\sigma_r$ and $\sigma_k$, $k = 1,\dots, 14$. The prior for $\nu_k$ and $\gamma_k$ are normal $N(0,1000)$ and normal $N(0,100)$, respectively. The SSVS models are fitted using the software JAGS \cite{plummer2003jags}, running 3 chains each with $200,000$ burn-in and $200,000$ samples with thinning $=10$. The other models are implemented in STAN \cite{carpenter2017stan}.

\subsection{Results} \label{subsec: resultReal}
The results are presented in Tables \ref{tab:DetroitInd} and \ref{tab:DetroitBeta}. Table \ref{tab:DetroitInd} shows the mean posterior probability of $I_k = 1$. For the Laplace shrinkage prior, we report whether 0 is outside of the $95\%$ credible intervals of the parameters. All SSVS models choose the same set of relevant outcomes in different setting. The informative prior on $p(I_k)$ results in different posterior mean of $I_k$, however, it does not affect the inference on the outcomes' relevance for most outcome variables.  The only exception is CBCL Externalizing at age 7, of which the posterior probability of being affected is slightly less than 0.5 (0.44 and 0.49) with a non-informative prior and slightly higher than 0.5 (0.53) with an informative prior. These results are also similar to the that of the Laplace shrinkage prior; however this prior shrinks more $\beta_k$ towards 0 than the SSVS priors. 

Table \ref{tab:DetroitBeta} presents the estimates of $\beta_k$ and overall effect $\mu$. The SSVS with informative $p(I_k = 1)$ and the subset model suggest a strong negative effect of PAE on the cognitive outcomes. These findings are consistent with those in \cite{jacobson2021effects}. The estimate of $\tau$ is similar for both models (0.19 vs 0.17). On the other hand, the SSVS models with the non-informative prior suggest a weaker effect (-0.33 and -0.32 versus -0.41). The non-informative prior also results in a larger estimate of $\tau$ (0.22 and 0.24 versus 0.19).   

Table \ref{tab:DetroitBeta} shows that all SSVS models produce smaller estimates for the coefficients of the affected outcomes and hence $\mu$, compared to the subset model. The result here is consistent with our observation in the simulation studies in Section \ref{sec:simExample}. However, as shown in Tables \ref{tab:DetroitInd} and \ref{tab:DetroitBeta}, our proposed model produces very similar estimates of $\beta_k$ compared to the Laplace prior in all settings. Table \ref{tab:DetroitBeta} also indicates that the informative prior for the indicator $I_k$ produces estimates of $\beta_k$ and $\mu$ that are closer to the subset model that only includes the affected outcomes.



\section{Conclusion} \label{sec:Conclusion}
In this paper we propose a statistical method for identifying outcomes from a large number of observed variables that are directly affected by an exposure variable. Our method is an extension of standard Bayesian variable selection models to multiple outcomes data, which also provides an estimate of the overall effect of the exposure variable. We demonstrate the performance and limitations of our method in two simulation exercises and a real data application.

Our application in modelling the effect of PAE on cognition identified a set of neurodevelopmental tests that are significantly affected by fetal alcohol exposure. In addition, the model indicates a negative overall effect of PAE on the sensitive outcomes. A limitation of the current model is that we only use an individual random intercept to capture the correlations among the outcomes. This approach may not be ideal, and we may consider a more sophisticated correlation structure in future works.

Finally, the proposed framework is shown to be effective in identifying sensitive outcomes in various scenarios. However, it may underestimate the outcome-specific effect size and mean effect when the effects are mild. This is a common issue with variable selection priors; we expect that the result can be improved by using more informative priors for the indicators $I_k$.

\section*{Acknowledgments}
This research was funded by grants from the National Institutes of Health/ National Institute on Alcohol Abuse and Alcoholism (NIH/NIAAA; R01-AA025905) and the Lycaki-Young Fund from the State of Michigan to Sandra W. Jacobson and Joseph L. Jacobson. Much of the work was accomplished while Khue-Dung Dang was a postdoctoral research fellow at UTS, supported by the Australian Research Council Centre of Excellence for Mathematical and Statistical Frontiers (ACEMS) grant CE140100049. Richard J. Cook was supported by the Natural Sciences and Engineering Research Council of Canada through grant RGPIN-2017-04207.

\section*{Data Availability Statement}
Original data used in the application available on request due to privacy/ethical restrictions. Standardized, de-identified data to replicate the results in the paper will be made available.
	\bibliographystyle{wileyNJD-AMA}
	\bibliography{ref}

\begin{thebibliography}{10}
\providecommand \doibase [0]{http://dx.doi.org/}%

\bibitem{hoyme2005practical}
Hoyme HE, May PA, Kalberg WO, et al. A practical clinical approach to diagnosis
  of fetal alcohol spectrum disorders: clarification of the 1996 institute of
  medicine criteria. {\it Pediatrics} 2005\string; 115(1)\string: 39--47.

\bibitem{hoyme2016updated}
Hoyme HE, Kalberg WO, Elliott AJ, et al. Updated clinical guidelines for
  diagnosing fetal alcohol spectrum disorders. {\it Pediatrics} 2016\string;
  138(2).

\bibitem{carter2016fetal}
Carter RC, Jacobson JL, Molteno CD, Dodge NC, Meintjes EM, Jacobson SW. Fetal
  alcohol growth restriction and cognitive impairment. {\it Pediatrics}
  2016\string; 138(2).

\bibitem{jacobson2004maternal}
Jacobson SW, Jacobson JL, Sokol RJ, Chiodo LM, Corobana R. Maternal age,
  alcohol abuse history, and quality of parenting as moderators of the effects
  of prenatal alcohol exposure on 7.5-year intellectual function. {\it
  Alcoholism: Clinical and Experimental Research} 2004\string; 28(11)\string:
  1732--1745.

\bibitem{jacobson2008impaired}
Jacobson SW, Stanton ME, Molteno CD, et al. Impaired eyeblink conditioning in
  children with fetal alcohol syndrome. {\it Alcoholism: Clinical and
  Experimental Research} 2008\string; 32(2)\string: 365--372.

\bibitem{mattson2019fetal}
Mattson SN, Bernes GA, Doyle LR. Fetal alcohol spectrum disorders: A review of
  the neurobehavioral deficits associated with prenatal alcohol exposure. {\it
  Alcoholism: Clinical and Experimental Research} 2019\string; 43(6)\string:
  1046--1062.

\bibitem{jacobson1993teratogenic}
Jacobson JL, Jacobson SW, Sokol RJ, Martier SS, Ager JW, Kaplan-Estrin MG.
  Teratogenic effects of alcohol on infant development. {\it Alcoholism:
  Clinical and Experimental Research} 1993\string; 17(1)\string: 174--183.

\bibitem{jacobson1999drinking}
Jacobson JL, Jacobson SW. Drinking moderately and pregnancy: effects on child
  development. {\it Alcohol research \& health} 1999\string; 23(1)\string:
  25-30.

\bibitem{jacobson2002effects}
Jacobson JL, Jacobson SW. Effects of prenatal alcohol exposure on child
  development. {\it Alcohol Research \& Health} 2002\string; 26(4)\string:
  282-286.

\bibitem{jacobson2021effects}
Jacobson JL, Akkaya-Hocagil T, Ryan LM, et al. Effects of prenatal alcohol
  exposure on cognitive and behavioral development: Findings from a
  hierarchical meta-analysis of data from six prospective longitudinal US
  cohorts. {\it Alcoholism: Clinical and Experimental Research} 2021.

\bibitem{lefkopoulou1993global}
Lefkopoulou M, Ryan L. Global tests for multiple binary outcomes. {\it
  Biometrics} 1993\string: 975--988.

\bibitem{budtz2002estimation}
Budtz-J{\o}rgensen E, Keiding N, Grandjean P, Weihe P. Estimation of health
  effects of prenatal methylmercury exposure using structural equation models.
  {\it Environmental Health} 2002\string; 1(1)\string: 1--22.

\bibitem{dunson2000bayesian}
Dunson DB. Bayesian latent variable models for clustered mixed outcomes. {\it
  Journal of the Royal Statistical Society: Series B (Statistical Methodology)}
  2000\string; 62(2)\string: 355--366.

\bibitem{sanchez2005structural}
S{\'a}nchez BN, Budtz-J{\o}rgensen E, Ryan LM, Hu H. Structural equation
  models: a review with applications to environmental epidemiology. {\it
  Journal of the American Statistical Association} 2005\string;
  100(472)\string: 1443--1455.

\bibitem{sammel2002effects}
Sammel MD, Ryan LM. Effects of covariance misspecification in a latent variable
  model for multiple outcomes. {\it Statistica Sinica} 2002\string: 1207--1222.

\bibitem{berkey1998meta}
Berkey C, Hoaglin D, Antczak-Bouckoms A, Mosteller F, Colditz G. Meta-analysis
  of multiple outcomes by regression with random effects. {\it Statistics in
  Medicine} 1998\string; 17(22)\string: 2537--2550.

\bibitem{van2015meta}
Noortgate V.~dW, L{\'o}pez-L{\'o}pez JA, Mar{\'\i}n-Mart{\'\i}nez F,
  S{\'a}nchez-Meca J. Meta-analysis of multiple outcomes: A multilevel
  approach. {\it Behavior Research Methods} 2015\string; 47(4)\string:
  1274--1294.

\bibitem{ryan2008combining}
Ryan L. Combining data from multiple sources, with applications to
  environmental risk assessment. {\it Statistics in Medicine} 2008\string;
  27(5)\string: 698--710.

\bibitem{liang1986longitudinal}
Liang KY, Zeger SL. Longitudinal data analysis using generalized linear models.
  {\it Biometrika} 1986\string; 73(1)\string: 13--22.

\bibitem{sammel1999multivariate}
Sammel M, Lin X, Ryan L. Multivariate linear mixed models for multiple
  outcomes. {\it Statistics in Medicine} 1999\string; 18(17-18)\string:
  2479--2492.

\bibitem{thurston2009bayesian}
Thurston SW, Ruppert D, Davidson PW. Bayesian models for multiple outcomes
  nested in domains. {\it Biometrics} 2009\string; 65(4)\string: 1078--1086.

\bibitem{turk2008analyzing}
Turk DC, Dworkin RH, McDermott MP, et al. Analyzing multiple endpoints in
  clinical trials of pain treatments: IMMPACT recommendations. {\it Pain}
  2008\string; 139(3)\string: 485--493.

\bibitem{akaike1998information}
Akaike H. Information theory and an extension of the maximum likelihood
  principle. In:  {\it Selected papers of Hirotugu Akaike} Springer.  1998 (pp.
  199--213).

\bibitem{schwarz1978estimating}
Schwarz G. Estimating the dimension of a model. {\it The Annals of Statistics}
  1978\string: 461--464.

\bibitem{hastie2020best}
Hastie T, Tibshirani R, Tibshirani R. Best Subset, Forward Stepwise or Lasso?
  Analysis and recommendations based on extensive comparisons. {\it Statistical
  Science} 2020\string; 35(4)\string: 579--592.

\bibitem{o2009review}
O'Hara RB, Sillanp{\"a}{\"a} MJ. A review of Bayesian variable selection
  methods: what, how and which. {\it Bayesian Analysis} 2009\string;
  4(1)\string: 85--117.

\bibitem{van2019shrinkage}
Van~Erp S, Oberski DL, Mulder J. Shrinkage priors for Bayesian penalized
  regression. {\it Journal of Mathematical Psychology} 2019\string; 89\string:
  31--50.

\bibitem{lefkopoulou1989analysis}
Lefkopoulou M, Moore D, Ryan L. The analysis of multiple correlated binary
  outcomes: Application to rodent teratology experiments. {\it Journal of the
  American Statistical Association} 1989\string; 84(407)\string: 810--815.

\bibitem{bondell2010joint}
Bondell HD, Krishna A, Ghosh SK. Joint variable selection for fixed and random
  effects in linear mixed-effects models. {\it Biometrics} 2010\string;
  66(4)\string: 1069--1077.

\bibitem{fan2012variable}
Fan Y, Li R. Variable selection in linear mixed effects models. {\it Annals of
  statistics} 2012\string; 40(4)\string: 2043.

\bibitem{yang2020bayesian}
Yang M, Wang M, Dong G. Bayesian variable selection for mixed effects model
  with shrinkage prior. {\it Computational Statistics} 2020\string;
  35(1)\string: 227--243.

\bibitem{george1993variable}
George EI, McCulloch RE. Variable selection via {G}ibbs sampling. {\it Journal
  of the American Statistical Association} 1993\string; 88(423)\string:
  881--889.

\bibitem{figueiredo2003adaptive}
Figueiredo MA. Adaptive sparseness for supervised learning. {\it IEEE
  Transactions on Pattern Analysis and Machine Intelligence} 2003\string;
  25(9)\string: 1150--1159.

\bibitem{park2008bayesian}
Park T, Casella G. The {B}ayesian {L}asso. {\it Journal of the American
  Statistical Association} 2008\string; 103(482)\string: 681--686.

\bibitem{carvalho2009handling}
Carvalho CM, Polson NG, Scott JG. Handling Sparsity via the Horseshoe. In:
  {\it Proceedings of the 12th International Conference on Artificial
  Intelligence and Statistics} PMLR; 2009\string: 73--80.

\bibitem{kass1995bayes}
Kass RE, Raftery AE. Bayes factors. {\it Journal of the American Statistical
  Association} 1995\string; 90(430)\string: 773--795.

\bibitem{spiegelhalter2002bayesian}
Spiegelhalter DJ, Best NG, Carlin BP, Linde AVD. Bayesian measures of model
  complexity and fit. {\it Journal of the Royal Statistical Society: Series B
  (Statistical Methodology)} 2002\string; 64(4)\string: 583--639.

\bibitem{watanabe2010asymptotic}
Watanabe S, Opper M. Asymptotic equivalence of {B}ayes cross validation and
  widely applicable information criterion in singular learning theory. {\it
  Journal of Machine Learning Research} 2010\string; 11(12).

\bibitem{kuo1998variable}
Kuo L, Mallick B. Variable selection for regression models. {\it Sankhy{\=a}:
  The Indian Journal of Statistics, Series B} 1998\string: 65--81.

\bibitem{dellaportas2002bayesian}
Dellaportas P, Forster JJ, Ntzoufras I. On Bayesian model and variable
  selection using MCMC. {\it Statistics and Computing} 2002\string;
  12(1)\string: 27--36.

\bibitem{meuwissen2004mapping}
Meuwissen TH, Goddard ME. Mapping multiple QTL using linkage disequilibrium and
  linkage analysis information and multitrait data. {\it Genetics Selection
  Evolution} 2004\string; 36(3)\string: 261--279.

\bibitem{plummer2003jags}
Plummer M. JAGS: A program for analysis of Bayesian graphical models using
  Gibbs sampling. In:  {\it Proceedings of the 3rd International Workshop on
  Distributed Statistical Computing} ; 2003\string: 1--10.

\bibitem{carpenter2017stan}
Carpenter B, Gelman A, Hoffman MD, et al. {STAN}: A probabilistic programming
  language. {\it Journal of Statistical Software} 2017\string; 76(1).

\bibitem{achenbach1991manual}
Achenbach TM. Manual for the Child Behavior Checklist/4-18 and 1991 profile.
  {\it University of Vermont, Department of Psychiatry} 1991.

\bibitem{wechsler1991manual}
Wechsler D. {\it Manual for the Wechsler Intelligence Scale for Children}.
\newblock Psychological Corporation.
\newblock 3~ed. 1991.

\bibitem{hocagil2020pscore}
Akkaya~Hocagil T, Cook RJ, Jacobson SW, Jacobson JL, Ryan LM. Propensity score
  analysis for a semi-continuous exposure variable: a study of gestational
  alcohol exposure and childhood cognition. {\it Journal of the Royal
  Statistical Society: Series A (Statistics in Society)} 2021\string;
  184(4)\string: 1390--1413.

\end{thebibliography}
	
\vspace{5em}
\begin{table}[!h]
\fontsize{10}{10}\selectfont
	\caption{The average number of outcomes correctly identified as relevant and incorrectly chosen as relevant in different settings with data generated from \eqref{eq:model_simple} (Study 1) and SEM (Study 2). The table shows the results from the original SSVS prior with $\mu = 0$ and our proposed prior where $\mu$ is unknown. $K_1$ is the true number of relevant outcomes. The fourth and fifth columns show the number of outcomes that each model detects as relevant. The next two columns show the number of relevant outcomes correctly identified by each model. The last two columns show the number of irrelevant outcomes that were detected as relevant. All numbers are averaged over 10 simulations.}
	\begin{tabular}{@{}l|l|c|cccccc@{}} \toprule
	Study &$\mu$	&	$K_1$&	\multicolumn{2}{c}{No. identified as relevant}&	\multicolumn{2}{c}{No. correctly identified} &	\multicolumn{2}{c}{No. incorrectly identified	} \\[5pt]
		 
&		 &	&$\mu$ unknown& 	$\mu = 0$ &$\mu$ unknown & 	$\mu = 0$ & $\mu$ unknown & 	$\mu = 0$ \\ [5pt]
		 \midrule
\multirow{6}{*}{$1$}&	\multirow{3}{*}{Small}
&5&	8.5&	5.5&	1.9&	1.3&	6.6&	4.2  \\ [5pt]
&&10&	6.6&	5.9&	3.2&	3.3&	3.4	&2.6  \\ [5pt]
&&15&	6.6	&5.8&	5.4&	4.2&	1.2&	1.6 \\ [5pt]
		
		\cline{2-9} \\[-8pt]
		
&	\multirow{3}{*}{Large}
&5&	5&	5&	5&	5&	0&	0\\ [5pt]
&&10&	10&	10&	10&	10&	0&	0\\ [5pt]
&&15	&15&	15&	15&	15&	0&	0\\ [5pt]
	
	\midrule
	
\multirow{6}{*}{$2$}&		\multirow{3}{*}{Small}	&5&	7.9&	7.1	&3.6&	3.2	&4.3&	3.9 \\ [5pt]
&&10&	9.3&	8&	8.2	&5.4&	1.1	&2.6 \\ [5pt]
&&15&	10.1&	7.3	&9.6&	4.6	&0.5&	2.7\\ [5pt]
		
			\cline{2-9} \\[-8pt]
			
&	\multirow{3}{*}{Large} &5	&5&	5&	5&	5&	0	&0  \\ [5pt]
&& 10&10&	10&	10&	10&	0&	0  \\ [5pt]
&& 15&15&	15&	15&	15&	0&	0 \\ [5pt]

		\bottomrule
	\end{tabular}	
	\vspace{3mm}
	\label{tab:simLong}
\end{table}

\begin{table}[!h]
	\fontsize{10}{10}\selectfont
	
	\caption{The mean squared errors of the models with different effect sizes, in simulated data study 1 and 2. The result is averaged over 10 simulations.}
	\begin{tabular}{@{}c|c|cccccc@{}} \toprule
	Study &	$K_1$&	\multicolumn{2}{c}{SSVS - $\mu$ unknown}&	\multicolumn{2}{c}{SSVS - $\mu= 0$} &	\multicolumn{2}{c}{No variable selection} \\[5pt]
		 
		& 	&Small $\mu$ & 	Large $\mu$ &Small $\mu$ & 	Large $\mu$ & Small $\mu$ & 	Large $\mu$ \\ [5pt]
		 \midrule
\multirow{3}{*}{1}&5&	0.008&	0.008&	0.002&	0.008&	0.027&	0.058  \\ [5pt]
&10&	0.009&	0.012&	0.006&	0.018&	0.018&	0.044  \\ [5pt]
&15&	0.009&	0.021&	0.008&	0.030&	0.014&	0.043 \\ [5pt]
		 \midrule
	\multirow{3}{*}{2}&5&	0.049&	0.023&	0.047&	0.020&	0.048&	0.060\\[5pt]
&10&	0.049&	0.031	&0.091&	0.046&	0.043&	0.068\\[5pt]
&15&	0.088&	0.035&	0.179&	0.053&	0.056&	0.066\\[5pt]
		\bottomrule
	\end{tabular}	
	\vspace{3mm}
	\label{tab:simMSE}
\end{table}

\begin{table}[!h]
	\fontsize{10}{10}\selectfont
	
	\caption{Estimates of $\mu$ in different settings of the simulated data study in Section \ref{subsec:simREdata}. The table shows the average of the posterior mean of $\mu$, averaged over 10 data sets, in different $\mu$ and $K_1$. The standard errors are in brackets. The first column is the true $\mu$. The subset model is the model that used only the correct set of relevant outcomes.}
	\begin{tabular}{@{}l|l|ccc@{}} \toprule
		$\mu$&	$K_1$	&	SSVS&	No selection&	Subset model  \\[5pt] \midrule
		\multirow{6}{*}{$-0.1$}
	
&5	&-0.021&	-0.003&	-0.073	\\[5pt]	
&	&(0.082)	&(0.161)&	(0.172)\\[5pt]	
&10	&-0.036	&-0.039&	-0.082 \\[5pt]	
&	&(0.068)	&(0.116)&	(0.116) \\[5pt]	
&15	&-0.061	&-0.114	&-0.142\\[5pt]	
&&	(0.085)&	(0.094)&	(0.107)	\\[5pt]

		\midrule
		\multirow{6}{*}{$-3$}
	
&5	&-2.906&	-0.703&	-2.903 \\[5pt]	
&&	(0.070)&	(0.175)&	(0.170)	 \\[5pt]	
&10	&-3.088&	-1.520&	-3.081 \\[5pt]	
&&	(0.047)	& (0.124)	&(0.122)	 \\[5pt]	
&15	&-2.948&	-2.237&	-2.962 \\[5pt]	
&&	(0.057)&	(0.110)&	(0.110)		\\[5pt]	
	
		\bottomrule
	\end{tabular}	
	\vspace{3mm}
	\label{tab:simMu}
\end{table}

\begin{table}[!h]
	\fontsize{10}{10}\selectfont
	
	\caption{Estimates of $\mu$ in different settings of the simulated data study in Section \ref{subsec:simSEMdata}. The table shows the average of the posterior mean of $\mu$, averaged over 10 data sets, for different $\omega$ and $K_1$. The standard errors are in brackets. The third column is the mean of the true loadings times $\omega$, which are considered as the true $\mu$. The subset model is the model that used only the correct set of relevant outcomes.}
	\begin{tabular}{@{}l|l|cccc@{}} \toprule
		$\omega$&	$K_1$&	Mean($\bm{\Lambda}_{1:K1,1}\times \omega$)	&	SSVS&	No selection&		Subset model  \\[5pt] \midrule
		\multirow{8}{*}{$-0.5$}
		&	5&	-0.498	&-0.178&	-0.098&	-0.470\\[5pt]
		&	& &(0.217)&	(0.136)&	(0.144)\\ [5pt]
		
		&	10 &	-0.5145&	-0.417	&-0.279	&-0.548 \\[5pt]
		& &	&(0.228)&	(0.110)&	(0.165) \\ [5pt]
		
		&	15 &	-0.509&	-0.268&	-0.296	&-0.399\\[5pt]
		&	&  &(0.123)&	(0.134)&	(0.138) \\ [5pt]

		\midrule
		\multirow{6}{*}{$-3$}&5	&-3.132&	-3.231&	-0.835&	-3.263  \\[5pt]
&&&	(0.224)&	(0.089)&	(0.201) \\[5pt]
&10&	-3.009	&-2.971&	-1.471&	-2.965 \\[5pt]
&&& 	(0.187)&	(0.126)&	(0.200) \\[5pt]
&15&	-3.144&	-3.160&	-2.318&	-3.105  \\[5pt]
&&&		(0.144)&	(0.154)&	(0.180)\\[5pt]	
		
		\bottomrule
	\end{tabular}	
	\vspace{3mm}
	\label{tab:simMu2}
\end{table}

\begin{table}[!h]
	\fontsize{9}{9}\selectfont
	\caption{Summary of $I_k$ for different models for Detroit data. For SSVS, the table shows the posterior mean of $I_k$. The table highlights the variables selected by the SSVS prior. For the other method, we report whether 0 is outside the $95\%$ credible interval of the corresponding $\beta_k$. The CBCL and TRF tests came from \cite{achenbach1991manual}; the IQ tests were based on the Wechsler Intelligence Test for Children-III \cite{wechsler1991manual}.  }
	\begin{tabular}{@{}lcccc @{}} \toprule
&Laplace prior	& \multicolumn{2}{c}{$p(I_k=1) = 0.5$} & Informative $p(I_k)$ \\ [5pt]	
	& &	$g_1 = \tau^2/100$ &	$g_1 = 0.2^2$ &	$g_1 = 0.2^2$ \\ [5pt] \midrule
CBCL Social Problem&	0&	0.285&	0.378&	0.370 \\ [5pt]
\textbf{CBCL Attention Problem}&	0&	0.509&	0.535&	0.580 \\ [5pt]
CBCL Internalizing&	0&	0.193&	0.240&	0.056 \\ [5pt]
CBCL Externalizing&	0&	0.438&	0.493&	0.532 \\ [5pt]
\textbf{TRF Social Problem}&	1&	0.860&	0.741&	0.948 \\ [5pt]
\textbf{TRF Attention Problem}&	1&	0.871&	0.742&	0.948 \\ [5pt]
TRF Internalizing&	0&	0.204&	0.259&	0.065 \\ [5pt]
\textbf{TRF Externalizing} &	1&	0.861&	0.747&	0.949 \\ [5pt]
Verbal IQ &	0&	0.293&	0.391&	0.386 \\ [5pt]
Performance IQ &	0&	0.340&	0.431&	0.444 \\ [5pt]
\textbf{Freedom from distractibility}&	1&	0.968&	0.813&	0.970  \\ [5pt]
\textbf{Verbal fluency}&	0&	0.671&	0.631&	0.900  \\ [5pt]
\textbf{Digit span backwards}	&1	&0.875&	0.738&	0.948  \\ [5pt]
Story memory&	0&	0.263&	0.353&	0.337 \\ [5pt]
		\bottomrule
	\end{tabular}
	\label{tab:DetroitInd}	
	\vspace{3mm}
\end{table}

\begin{table}[!h]
	\fontsize{9}{9}\selectfont
	\caption{Posterior mean of $\beta_k$ and $\mu$ for Detroit data. For SSVS, we report the mean of $\beta_k|I_k = 1$ if the posterior mean of $I_k$ is greater than $0.5$ and $0$ otherwise. }
	\begin{tabular}{@{}lccccc @{}} \toprule
&Laplace prior	& \multicolumn{2}{c}{$p(I_k=1) = 0.5$} & Informative $p(I_k)$ & Subset\\ [5pt]	
	& &	$g_1 = \tau^2/100$ &	$g_1 = 0.2^2$ &	$g_1 = 0.2^2$ &  \\ [5pt] \midrule
CBCL Social Problem	&-0.089	&0.000&	0.000	&0.000&	\\ [5pt]
\textbf{CBCL Attention Problem}	&-0.226	&-0.274&	-0.275	&-0.346	&-0.493 \\ [5pt]
CBCL Internalizing &0.086&	0.000&	0.000&	0.000&	\\ [5pt]
CBCL Externalizing &	-0.194&	0.000&	0.000&	-0.330&	\\ [5pt]
\textbf{TRF Social Problem} &	-0.482&	-0.406	&-0.419	&-0.464	&0.629 \\ [5pt]
\textbf{TRF Attention Problem}  &	-0.459 &	-0.396&	-0.411	&-0.456&	-0.621 \\ [5pt]
TRF Internalizing&	0.065&	0.000&	0.000&	0.000&	\\ [5pt]
\textbf{TRF Externalizing} &	-0.498&	-0.409&	-0.423&	-0.469&	-0.636 \\ [5pt]
Verbal IQ &-0.111	&0.000&	0.000&	0.000	&\\ [5pt]
Performance IQ &-0.137&	0.000&	0.000& 0.000&	\\ [5pt]
\textbf{Freedom from distractibility}&	-0.560	&-0.447	&-0.471	&-0.509&	-0.640 \\ [5pt]
\textbf{Verbal fluency}&	-0.325	&-0.332&	-0.336&	-0.393&	-0.543 \\ [5pt]
\textbf{Digit span backwards }&	-0.452&	-0.391	&-0.405&	-0.452&	-0.591 \\ [5pt]
Story memory&	-0.057&	0.000&	0.000&	0.000&	 \\ [5pt]
$\mu$	&	&-0.326&	-0.312&	-0.405&	-0.594 \\ [5pt]
$\tau$	& &	0.216&	0.241&	0.187&	0.171\\ [5pt]
		\bottomrule
	\end{tabular}
	\label{tab:DetroitBeta}	
	\vspace{3mm}
\end{table}
\end{document}